\documentclass[prb,showpacs,preprintnumbers,amsfonts,amssymb,amsmath,floats,twocolumn,superscriptaddress,aps]{revtex4}

\usepackage{graphicx}
\usepackage{dcolumn}
\usepackage{bm}
\usepackage[final,dvips]{epsfig}
\usepackage{bm}
\usepackage{color}

\newcommand{\ocite}[1]{Ref.\ \onlinecite{#1}}
\newcommand{\oocite}[1]{Refs.\ \onlinecite{#1}}
\newcommand{\ff}[1]{{\bm #1}}
 
\begin{document} 

\title{
A Dynamical Quantum Cluster Approach to
Two-Particle Correlation \\ Functions in the Hubbard Model
} 

\author{S.~Hochkeppel}

\author{F.~F.~Assaad}

\affiliation{
Institute for Theoretical Physics, University of
W\"urzburg, 97074 W\"urzburg, Germany
}

\author{W.~Hanke}

\affiliation{
Institute for Theoretical Physics, University of
W\"urzburg, 97074 W\"urzburg, Germany
}
\affiliation{
Kavli Institute for Theoretical Physics, University of California,
Santa Barbara, CA 93106, USA
}
 
\begin{abstract} 
We investigate the charge- and spin dynamical structure factors for the 2D
one-band Hubbard model in the strong coupling regime within an
extension of the Dynamical
Cluster Approximation (DCA) to two-particle response functions.  The
full  irreducible two-particle vertex with three momenta and
frequencies  is approximated by 
an effective vertex  dependent on the momentum  and frequency of the
spin/charge excitation. In the spirit of the DCA,
the effective vertex is calculated with quantum Monte Carlo methods on a
finite cluster.  On the basis of a comparison with high temperature auxiliary field quantum
Monte Carlo  data   we show that near and beyond optimal doping, our results provide a consistent
overall picture of the interplay between charge,  spin and single-particle
excitations. 
\end{abstract} 

\maketitle  


\section{Introduction}
\label{sec:intro}
Two-particle correlation functions, such as the  dynamical spin- and
charge correlation functions, determine a variety of crucial properties
of many-body systems. Their poles as a function of frequency and
momentum describe the elementary excitations, i.e. electron-hole
excitations and collective modes, such as spin- and charge-density
waves. Furthermore, an effective way to identify continuous phase
transitions is to search for divergencies of
susceptibilities, i.e. two-particle correlation functions. Yet,
compared to studies of single-particle Green's functions and their
spectral properties, where a good overall accord between theoretical
models (Hubbard type-models) and experiment (ARPES) has been
established (see
\oocite{DHS2003,preuss95,preuss97,IFT98}), the situation
is usually not so satisfying for 
two-particle Green's functions. This is especially so for the case of
correlated electron systems such as high-$T_c$ superconductors
(HTSC). The primary reason for this is that calculations of these
Green's functions are, from a numerical point of view, much more
involved. 

\begin{figure}[t]
\centering
\includegraphics[width=0.8\columnwidth]{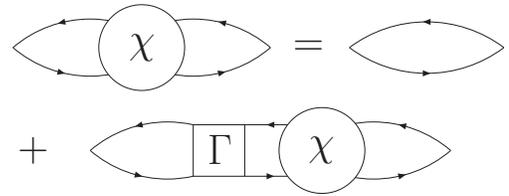} 
\caption{ Bethe-Salpeter equation for the two-particle propagator.}
\label{bethe_salp_pic}
\end{figure}

To expose the problem let us  consider the spin-susceptibility  which is
given by: 
\begin{eqnarray}
  \chi( \underline{q} ) =   \frac{1}{\beta L }  \sum_{ \underline{k}, \underline{p} } \chi_{   \underline{k}, \underline{p} } 
    ( \underline{q} )  \; \; {\rm ,with}  \\
   \chi_{   \underline{k}, \underline{p} }  ( \underline{q} ) =  
 \langle c^{\dagger}_{\underline{k},\uparrow} c^{\phantom{\dagger}}_{\underline{k}+  \underline{q},\downarrow} 
        c^{\dagger}_{\underline{p},\downarrow} c^{\phantom{\dagger}}_{\underline{p}-  \underline{q},\uparrow}  \rangle 
\nonumber
\end{eqnarray}  
Here,  $L$ corresponds  to the lattice size, $\beta $ is the inverse temperature and  $\underline{q} \equiv ( \ff q, \Omega_m) $, 
$\ff q$  being the momentum and $ \Omega_m $  a (bosonic) Matsubara  
frequency.  To simplify the notation, we have adopted a path integral coherent state notation with Grassman variables: 
\begin{equation}
  c_{\underline{k},\sigma } \equiv c_{\ff k, \omega_m,\sigma} = \frac{1}{\sqrt {\beta L} } \sum_{\ff r} \int_{0}^{\beta} {\rm d } \tau 
         e^{i (\omega_m \tau - \ff k \ff r)} c_{ \ff r,\sigma } (\tau)
\end{equation}
The two-particle irreducible vertex, $ \Gamma_{\underline{k}', \underline{k}'' } (\underline{q} ) $,  is defined through 
the Bethe-Salpeter equation,
\begin{equation}
\label{bethe_salp}
   \chi_{\underline{k}, \underline{p} }  ( \underline{q} ) = \chi^{0}_{\underline{k}, \underline{p} } (\underline{q} )  
+   \sum_{\underline{k}', \underline{k}'' } 
 \chi^{0}_{\underline{k}, \underline{k}' } (\underline{q} )  \Gamma_{\underline{k}', \underline{k}'' } (\underline{q} ) 
 \chi_{\underline{k}'', \underline{p} }  ( \underline{q} ), 
\end{equation}
which is diagrammatically depicted in Fig. \ref{bethe_salp_pic}.

Within the  Dynamical Cluster Approximation (DCA)~\cite{HTZ+98,MJPH05},  which we
consider in this paper,  one can consistently define the two-particle Green's
functions, by extracting the irreducible vertex function from the
cluster. 
This approximation  maps the original lattice problem to a cluster of size  $L_c=l_c \times l_c$ 
embedded in a self-consistent host.  Thus, correlations up to a range $\xi < l_c $
are treated accurately, while the physics at longer  length-scales  is described at the mean-field level.
The DCA is conveniently formulated in momentum space, i.e. via a
patching of the BZ. Let ${ \ff K }$ denote such a patch, and 
${\ff k}$ the original lattice momentum.  The approximation boils down to
restricting momentum conservation only to the patch wave vectors 
${\ff K}$.  This approximation is justified if $k$-space correlation
functions are rather structureless and, thus, in real space short-ranged. 
From the technical point of view, the approximations are implemented via
the Laue function $ \Delta( {\ff k_1}, {\ff k_2}, {\ff k_3}, {\ff k_4} )$, 
which guarantees momentum conservation up to a reciprocal lattice vector. In 
the DCA, the Laue function  is replaced
by  $ \Delta_{DCA}( {\ff K_1}, {\ff K_2}, {\ff K_3}, {\ff K_4} ) $,
thereby insuring momentum conservation only between the cluster
momenta.  It is understood 
that   ${\ff k}_i$ belongs to the  patch  ${\ff K}_i$. 

To define uniquely the 
DCA approximation, in particular in view of two-particle quantities, it is useful 
to start with the Luttinger-Ward functional $\Phi$, which is computed using the DCA Laue function.    
Hence, $\Phi_{DCA} $ is a functional of a coarse-grained Green's function,  
$\bar{G}({\ff K},i \omega_m) \equiv \bar{G}(\underline{K} ) $.  Irreducible quantities such as the self-energy, 
and the  two-particle 
irreducible vertex are calculated on the 
cluster and correspond, respectively, to the first- and second-order functional derivatives
of  $\Phi_{DCA} $ with respect to $\bar{G}$.  
Using the cluster irreducible self-energy, $\Sigma({\underline{K}})$, and two-particle vertex, 
$ \Gamma_{\underline{K}',\underline{K}''}(\underline{Q})$,  one can then 
compute the lattice single-particle and lattice two-particle correlation functions using the 
Dyson and Bethe-Salpeter equations. This construction of two-particle quantities has 
the appealing property that  they are thermodynamically consistent\cite{Bay62,BK61}. Hence, 
the spin susceptibility  as calculated using the particle-hole correlation functions
corresponds precisely to the derivative of the magnetization with respect to an applied 
uniform static  magnetic field. 
The technical aspects of the above  program  are readily carried out for single-particle
properties. 
However a full calculation of the  irreducible two-particle vertex --- even
within the DCA  --- 
is prohibitively  expensive~\cite{JMHM01} and, thus, has never been carried
out.

In the present work, we would like to overcome this situation by
suggesting a scheme where the $\underline{K}'-$ and $\underline{K}''-$
dependencies of the irreducible vertex are neglected. At low temperatures, 
this amounts  to the assumption that  in a energy and momentum window around 
the Fermi surface, the irreducible vertex depends weakly on $\underline{K}'-$ and $\underline{K}''$.     
Following this assumption, an effective two-particle vertex  in terms of an average over
the $\underline{K}'-$ and $\underline{K}''$ dependencies of $\Gamma_{\underline{K}',\underline{K}''}(\underline{Q})$ 
is introduced: 
\begin{equation}
	\frac{1}{\beta L} U_{eff}(\underline{Q})  = \langle \Gamma_{\underline{K}',\underline{K}''}(\underline{Q}) \rangle.
\label{gamma_aver}
\end{equation}
As shown in an earlier Quantum Monte Carlo (QMC) study by Bulut et
al. (\ocite{BSW93}) for a single QMC cluster, this is reasonable 
for the 2D Hubbard model (on this QMC cluster of size $8 \times 8$
with $U = 8t$).
The authors of \ocite{BSW93} have also calculated
with $U_{eff}(\underline{Q} )$ the effective electron-electron
interaction for the $8 \times 8$ single QMC cluster.  Both the momentum and frequency
dependence were in rather good agreement with the QMC results for the effective electron-electron interaction.

Replacing  the irreducible vertex by $\frac{1}{\beta L} U_{eff}(\underline{Q}) $  in the cluster version of the 
Bethe-Salpeter Eq. (\ref{bethe_salp}) and carrying out the summations to obtain the cluster susceptibility  gives: 
\begin{equation}
  U_{eff}(\underline{ Q}) = \frac{1}{\bar{\chi}_0(\underline{ Q} ) } - 
  \frac{1}{\chi(\underline{ Q}) }, 
\label{ueff} 
\end{equation}
where $\chi$ corresponds to the fully interacting susceptibility on the DCA cluster
in the particle-hole channel and $\bar{\chi}_0$ is the corresponding bubble as 
obtained from the coarse-grained Green's functions.
Finally, our estimate of  the lattice susceptibility reads: 
\begin{equation}
  \chi( \underline{q} ) = \frac{\chi_0(\underline{q})} 
  {  1 - U_{eff} (\underline{Q}) \cdot \chi_0(\underline{q}) }.
\label{rpa-eq}
\end{equation} 
where $\chi_0(\underline{q}) $ corresponds to the bubble of the dressed  lattice Green's functions: 
$G(\underline{k}) = \frac{1}{ G_0^{-1}(\underline{k}) - \Sigma(\underline{K})} $.

In the following section, we describe some aspects of the explicit implementation of the DCA which is 
based on the Hirsch-Fye QMC algorithm~\cite{HF86}.  Our new approach requires substantial testing. In Sec. 
\ref{subsec:resA} we compare the  N\'eel temperature as obtained  within the DCA without any further 
approximations to the result obtained with  our new approach.  In Sec. \ref{subsec:resB}, we present results for the 
temperature and doping dependence of the  spin and charge dynamical  structure factors and compare the high temperature 
data with auxiliary field QMC simulations.  Finally,  Sec. \ref{sec:con} draws conclusions.

\section{Implementation of the DCA}
\label{sec:th}

We consider the standard model of strongly correlated electron systems, the single-band Hubbard model \cite{Hub63,Gut63,Kan63}
\begin{equation}
  H = - \sum_{ij \sigma} t_{ij} c_{i\sigma}^\dagger c_{j\sigma} 
  + U \sum_i n_{i\uparrow} n_{i\downarrow},
\end{equation}
with hopping between nearest neighbors $t$ and Hubbard interaction $U$. The energy scale is set  by
$t$ and 
throughout the paper we consider $U=8t$.

Our goal is to compute the  spin, $S(\ff q,\omega)$   and charge  $C(\ff q,\omega)$   dynamical structure factors. They are given, 
respectively, by:
\begin{subequations}
\begin{align}
 & &  \langle S^z(\ff q, \tau)  S^z(-\ff q,0) \rangle  = \frac{1}{\pi} \int {\rm d} w \; e^{-\tau \omega} \; S(\ff q,\omega)  \label{second} \\ 
 & &  \langle N(\ff q, \tau)   N(-\ff q,0) \rangle  = \frac{1}{\pi}
 \int {\rm d} w \; e^{-\tau \omega} \; C(\ff q,\omega) \label{third}.
\end{align}
\end{subequations}
Here, $ S^z(\ff q)  = \frac{1}{\sqrt {L} } \sum_{\ff j} e^{i \ff q \ff  j } \left( n_{j,\uparrow} - n_{j,\downarrow}  \right) $  and 
 $ N(\ff q)  = \frac{1}{\sqrt {L} } \sum_{\ff j} e^{i \ff q \ff  j } \left( n_{j,\uparrow} + n_{j,\downarrow}  \right) $ . 
The left hand side of the above equations are obtained from the corresponding susceptibility as calculated from Eq. (\ref{rpa-eq}). 
Finally, a stochastic version of the Maximum Entropy method \cite{Sandvik98,beach-2004} is used to extract the 
dynamical quantities.

In order to cross check our results,  we slightly modify  Eq. (\ref{rpa-eq})  to: 
\begin{equation}
  \chi( \underline{q}) = \frac{\chi_0(\underline{q})} 
  {  1 - \alpha \cdot U_{eff} (\underline{q} ) \cdot \chi_0(\underline{q}) }.
\label{rpa-eq-alpha}
\end{equation}
Here, we have introduced an  additional "controlling" parameter $\alpha$   in the susceptibility denominator, 
which is calculated in a self-consistent manner. It assures, for example in the case of the longitudinal spin response, that
$\chi( \underline{q} )$ obeys the following sum rule (a similar idea,
to use sumrules for constructing a controlled local approximation for
the irreducible two-particle vertex has been implemented by Vilk and
Tremblay (\ocite{vilk97})):
\begin{equation}
  \frac{1}{\beta L}  \sum_{ \underline{q}} \chi( \underline{q} ) = \langle (S^z_i)^2 \rangle.
\label{sumrule}
\end{equation}
Of course, $\alpha$ should be as close as possible equal to $\alpha =1$, which is indeed what we will find after implementing 
the sum rule (see below).

Our implementation of the DCA  for the Hubbard model is standard \cite{MJPH05}. Here, we will only discuss our interpolation 
scheme as well as the implementation of a SU(2)-spin symmetry broken algorithm. 
Since the DCA  evaluates the  irreducible quantities, $ \Sigma(\underline{K})$ as well as $U_{eff}(\underline{Q})$  for 
the cluster wave vectors, an interpolation scheme has to be used. To this, we adopt the following 
strategy:  for a fixed Matsubara frequency $i \Omega_m$ and for each cluster vector
$\ff Q$, the effective interaction $U_{eff}$ is rewritten as a series 
expansion: 
\begin{equation}
U_{eff}(\ff Q, i \Omega_m) = \sum_i \sum_{\Delta_i} e^{i\Delta_i\ff Q} A_i(i \Omega_m),
\label{expansion}
\end{equation} 
with $i=0,...,n-1$, where $n$ is the number of the cluster
momentum vectors $\ff Q$. The quantity $\Delta_i$ represents vectors, where each vector from the corresponding
$\Delta_i$ belongs to the same "shell" around the origin $(0,0)$ in real
space, i.e.
\begin{equation}
  \Delta_0 = \left(\begin{array}[c] {c}  0 \\ 0 \end{array} \right);
     \Delta_1 = \pm \left(\begin{array}[c] {c}  1 \\ 0 \end{array} \right),
  	\pm \left(\begin{array}[c] {c}  0 \\ 1 \end{array} \right) \nonumber
\end{equation}
\begin{equation}
  \Delta_2 = \left(\begin{array}[c] {c}  \pm 1 \\ \pm 1 \end{array}
  \right),  \left(\begin{array}[c] {c}  \mp 1 \\ \pm 1
    \end{array}\right); 
  \Delta_3 = \pm \left(\begin{array}[c] {c}  2 \\ 0 \end{array} \right),
  	\pm \left(\begin{array}[c] {c}  0 \\ 2 \end{array} \right)
        ... \;. \nonumber
\label{delta}
\end{equation}
\begin{figure}[t]
\centering
\includegraphics[width=\columnwidth]{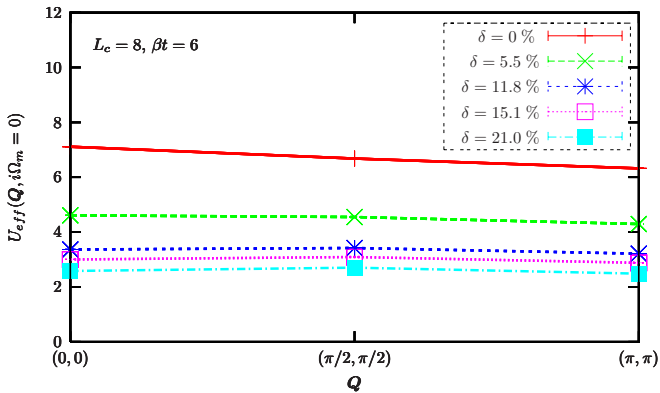}
\caption{Static $(i\Omega_m = 0)$ irreducible particle-hole interaction $U_{eff}$ for
  different cluster momentum vectors and dopings at inverse
  temperature $\beta t = 6$ and $U = 8t$. 
}
\label{ueff_k.fig}  
\end{figure} 
With a given $U_{eff}$, Eq. (\ref{expansion}) can be inverted to uniquely determine
$A_i$. With these coefficients, one can compute the effective particle-hole interaction for every
lattice momentum vector $\ff{q}$.  This interpolation method works
well when $U_{eff}$ is localized in real space and the sum in
Eq. (\ref{expansion}) can be cut-off at a given shell. 

The effective particle-hole interaction $U_{eff}$ is shown in
Fig. \ref{ueff_k.fig} for a variety of dopings at inverse temperature
$\beta t = 6$, $U/t = 8$ and on an $L_c = 8$ cluster, which corresponds
to the so-called "8A" Betts
cluster (see \oocite{BLF99,MJSKW05}).  The $U_{eff}$-function displays a 
smooth momentum dependence. These observations further support the 
interpolation scheme (Eq. \ref{expansion}). Thus,
indeed, $U_{eff}$ is rather localized in real space with sizable
reduction from its bare $U =8t$ value for larger doping and a further 
slight reduction at $\ff q = (\pi,\pi)$. The reduction is partly due
to the self-energy effects in the single-particle propagator, which
reduce $\bar{\chi}_0$ from its non-interacting ($U=0$) value
$\chi^{(0)}$. Partly, it also reflects both the Kanamori 
(see \ocite{Kan63}) repeated particle-particle scattering and vertex
corrections. 

\begin{figure}[t]
\centering
\includegraphics[width=0.8\columnwidth]{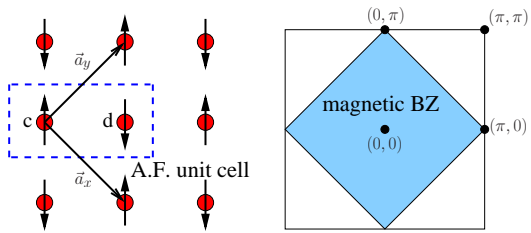}
\caption{SU(2) symmetry broken DCA calculation. Left: A.F. unit cell
  with new basis vectors in realspace. Right: Reduced A.F. unit cell
  in momentum space. 
}
\label{su2.fig}   
\end{figure}
Summarizing, the new approach to two-particle properties relies on  two approximations
which render the calculation of the corresponding Green's function possible. Firstly, 
the effective particle-hole interaction $U_{eff}(\underline{Q})$ depends only on
the center-of-mass momentum and frequency, i.e. $\ff Q$ and
$i\Omega_m$.  Secondly,  $\chi(\underline{Q})$, is extracted
directly from the cluster and  $\bar{\chi}_0( \underline{Q} )$  is
obtained  from the bubble of the coarse-grained Green's functions.

To generate  DCA results for the N\'eel temperature, we have used an SU(2) symmetry broken code. 
The  setup  is illustrated in
Fig. \ref{su2.fig}. We introduce a doubling of the unit cell --- to accommodate
AF  ordering --- which in turn defines the magnetic Brillouin zone.  The DCA k-space patching
is carried out in the magnetic Brillouin zone and  
the Dyson equation for the
single-particle propagator is given as a matrix equation:
\begin{equation}
G^{\sigma}(\underline{k}) =  \frac{1}{{G_0}^{-1}(\underline{k} ) - \Sigma^\sigma(\underline{K})},
\label{self-eq}
\end{equation}
with
\begin{equation}
G^{\sigma}(\underline{k}) = \left( \begin{array}[c]{cc}
  G_{cc}^{\sigma}(\underline{k})  & G_{cd}^{\sigma}(\underline{k})  \\
  G_{dc}^{\sigma}(\underline{k})  & G_{dd}^{\sigma}(\underline{k})  \end{array} \right).
\label{self-matrix}
\end{equation}
With the SU(2) symmetry broken  algorithm, one can compute directly the staggered  magnetization,
i.e. $m = \frac{1}{L} \sum_{\ff j} e^{i \ff Q \ff j }(n_{\ff j,\uparrow} - n_{\ff j,\downarrow})$,  
and thereby  determine the transition temperature. 
Since the DCA is a conserving approximation, the so determined  transition temperature corresponds  precisely  to 
the temperature scale at which the corresponding susceptibility, calculated without any approximations on the 
irreducible vertex $\Gamma_{\underline{K}', \underline{K}''}(\underline{Q}) $, diverges.

\section{Results}
\subsection{Comparison of the AF transition temperature with 
a symmetry broken  DCA calculation}
\label{subsec:resA}

\begin{figure}[t]
\centering
\includegraphics[width=\columnwidth]{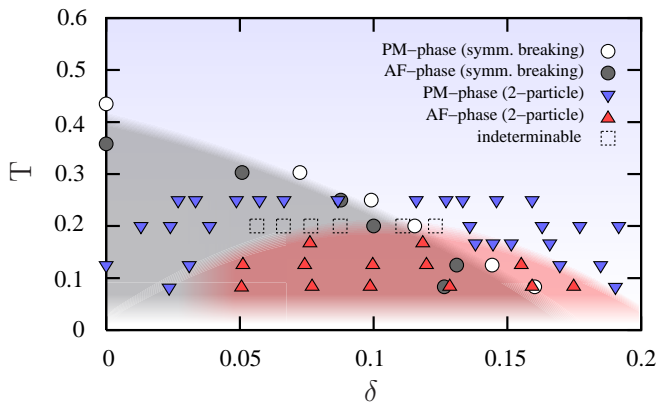}
\caption{Phase diagram for the one-band Hubbard model with $U = 8t$
  with respect to  different temperatures and fillings. The calculations are
  carried out on an $L_c=8$ cluster. The red and gray (blue and blank)
  objects indicate the antiferromagnetic (paramagnetic) phase. Shading
  of AF and PM regions is a guide to the eye. Details are in the text.
}
\label{magnetisation.fig}   
\end{figure}

A first test of the validity of our new approach is a comparison with the  
SU(2) symmetry broken DCA calculation on an $L_c = 8$ cluster at $U = 8t$.
The idea is to extract the N\'eel temperature $T_N$  from a divergence in the 
spin susceptibility as calculated in the above described (paramagnetic)
scheme --- see Eq. (\ref{rpa-eq-alpha}) ---  and to compare it  to  the DCA result as obtained from  the 
SU(2) symmetry broken algorithm.     This comparison provides information on the accuracy of our approximation 
to the  two-particle irreducible vertex (see Eq.~\ref{gamma_aver}).

Using the SU(2) symmetry breaking algorithm,  the magnetic phase diagram
for the  one-band Hubbard model as a function of doping is shown in
Fig. \ref{magnetisation.fig}.   
The para-(antiferro)magnetic phase transition
is indicated here by gray (blank) circles.  At half-filling  $T_N \simeq 0.4 t $ and  magnetism survives 
up to  approximately 15 \% hole doping.   It is know that the convergence of the
magnetization during the self-consistent steps in the DCA approach
is extremely poor near the phase transition and, therefore, we cannot
estimate the transition temperature more precisely than shown in Fig. \ref{magnetisation.fig}.
However, our precision is sufficient for comparison.  
We again stress that  the  so determined  magnetic phase diagram corresponds to
the {\it exact } DCA result where no approximation --- apart from
coarse graining --- is made on the  particle-hole irreducible vertex.

\begin{figure}[t]
\centering
\includegraphics[width=\columnwidth]{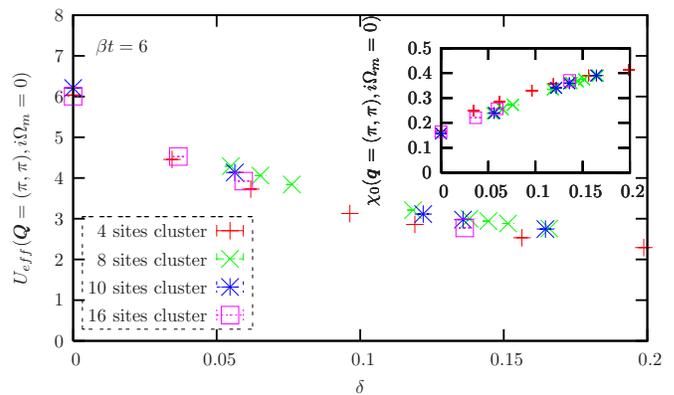}
\caption{Static irreducible particle-hole interaction $U_{eff}$ for
  the cluster momentum vector $\ff Q = (\pi,\pi)$. The inset shows the
  static free lattice susceptibility $\chi_0$ for the momentum vector $\ff q =
  (\pi,\pi)$. The bare Hubbard interaction strength is $U = 8t$.
}
\label{screening.fig}   
\end{figure}

The blue (red) triangles indicate the transition line for the para- to the antiferromagnetic solutions extracted
from the divergent spin susceptibility (Eq. \ref{rpa-eq-alpha}) within the
paramagnetic calculation. A precise estimation of the N¸\'eel  temperature  requires very accurate results  and boils down 
to finding the  zeros of the denominator of Eq. (\ref{rpa-eq-alpha}). 
In Fig.  \ref{screening.fig},  we consider the 
effective irreducible particle-hole interaction $U_{eff}$ for the
static case  and for the cluster momentum $\ff Q = (\pi,\pi)$ relevant for the AF instability. 
As apparent,  the irreducible particle-hole interaction  becomes
weaker  with increasing doping. On the other hand, the
susceptibility $\chi_0({\ff q},i\Omega_m=0)$ grows with
increasing doping.  
At a first glance both quantities $U_{eff}$ and $\chi_0$ 
(see Fig. \ref{screening.fig}) vary  smoothly as a function of
doping. However, in the vicinity of the phase transition, signalized
by  the  vanishing  of the  denominator in Eq. (\ref{rpa-eq-alpha}),  
the precise interplay between $U_{eff}$ and $\chi_0$
becomes delicately important and renders an accurate estimate of the N\'{e}el 
temperature difficult.   
Given the difficulty in determining precisely the N\'eel temperature, we obtain  good agreement 
between both methods at   $\delta \gtrsim 10 $ \%.
Note that in those calculations the values of $\alpha \approx 0.86 - 0.97$ are required  to  satisfy the sum rule in 
Eq. (\ref{sumrule}).   
At smaller dopings, and in particular at half-band filling the N\'eel temperature, as determined by  the vanishing
of the denominator in Eq. (\ref{rpa-eq-alpha}),  
underestimates the DCA result.  Hence, in this limit, the $\underline{K}' $ and $\underline{K}'' $  dependence 
of the irreducible vertex plays an important role in the determination of $T_N$ and cannot be neglected. 

Let us  emphasize, that a good agreement between the N\'eel temperatures at $\delta \simeq 10$ \% and above 
is a non-trivial achievement lending substantial support to the above new scheme for extracting
two-particle  Green's functions. 

\subsection{Dynamical Spin and Charge structure factors}

\label{subsec:resB}

\begin{figure}[t]
  \begin{minipage}[t]{0.475\columnwidth}
    \flushleft{a)}
    \vfill
    \centering
    \resizebox{1.0\columnwidth}{!}{\includegraphics{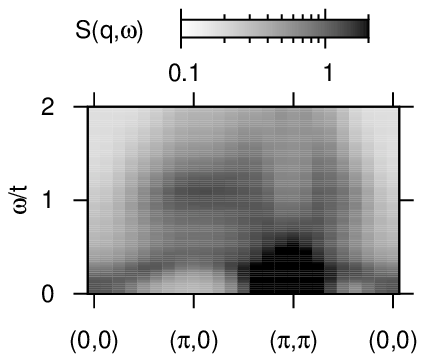}}
  \end{minipage}
  \hfill
  \begin{minipage}[t]{0.475\columnwidth}
    \flushleft{b)}
    \vfill
    \centering
    \resizebox{1.0\columnwidth}{!}{\includegraphics{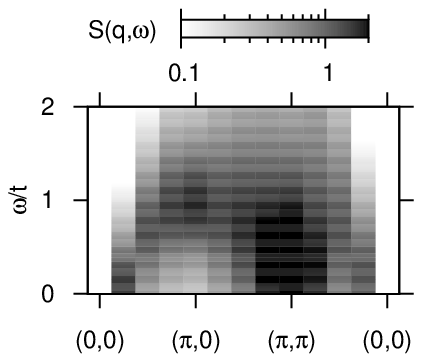}}
  \end{minipage}
  \begin{minipage}[t]{0.475\columnwidth}
    \flushleft{c)}
    \vfill
    \centering
    \resizebox{1.0\columnwidth}{!}{\includegraphics{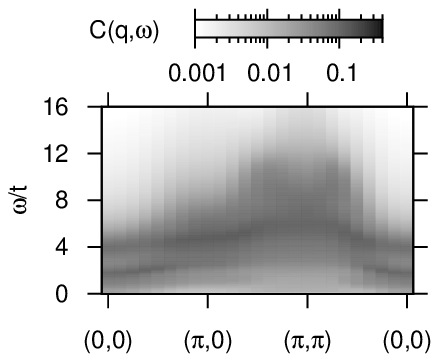}}
  \end{minipage}
  \hfill
  \begin{minipage}[t]{0.475\columnwidth}
    \flushleft{d)}
    \vfill
    \centering
    \resizebox{1.0\columnwidth}{!}{\includegraphics{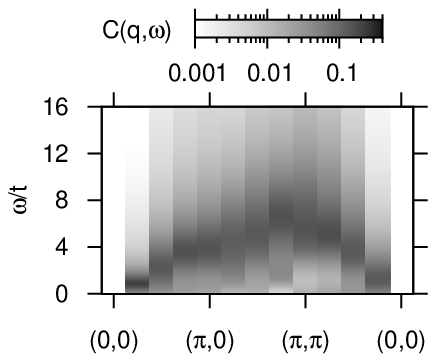}}
  \end{minipage}
  \caption{DCA (left) versus auxiliary field QMC (BSS) (right) for the dynamical spin
    and charge structure factors of the Hubbard model at $U/t=8$, $\delta \approx
    14 \; \%$ and $\beta t = 3$. The BSS data on the 
    $8\times 8$ lattice is essentially exact and acts as a benchmark
    for the DCA approach. The DCA calculations were carried out on an
    $L_c=8$ cluster.  Here we have used $\alpha = 0.98$  and $\alpha = 1.01$ to
    satisfy the sum rule in the spin and charge sectors respectively. 
  }
  \label{DCA_BSS.fig} 
\end{figure}

To further asses the validity of 
our approach, we compare  it  to {\it exact}  auxiliary-field  Blankenbecler,
Scalapino,  Sugar (BSS) QMC results (\ocite{preuss97}).  This
method  has a severe sign-problem especially in the vicinity of $\delta \simeq 10$ \%
and, hence, is restricted to high temperatures. 

\begin{figure}[t]
  \begin{minipage}[t]{\columnwidth}
    \flushleft{a)}
    \vfill
    \centering
    \resizebox{0.85\columnwidth}{!}{\includegraphics{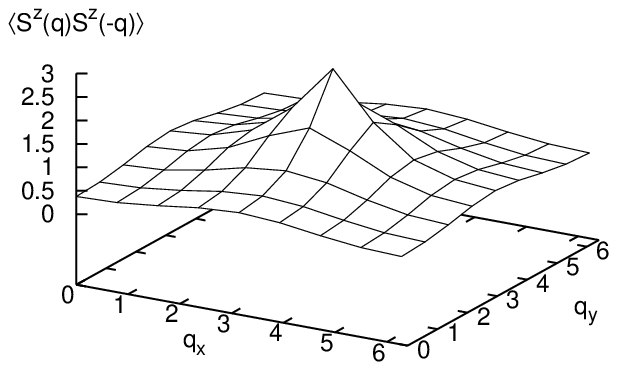}}
  \end{minipage}
  \begin{minipage}[t]{\columnwidth}
    \flushleft{b)}
    \vfill
    \centering
    \resizebox{0.85\columnwidth}{!}{\includegraphics{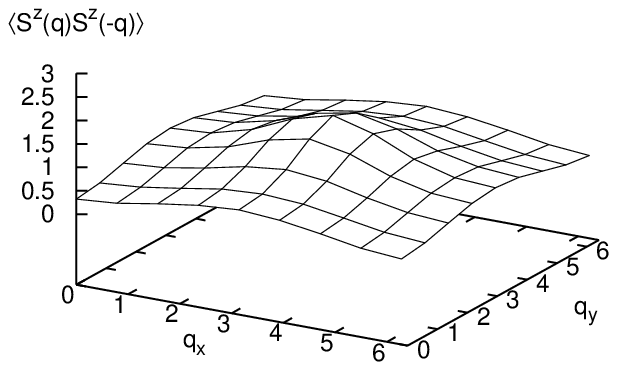}}
  \end{minipage}
\caption{DCA a) versus BSS b) static spin correlation function at 
   $U/t=8$, $\delta \approx 14 \; \%$ and $\beta t = 3$ on an
   $L_c=8$ cluster.
 }
 \label{static.spin.fig}   
 \end{figure}

Such a comparison for the dynamical spin,  $S(\ff q,\omega)$,  and
charge, $ C(\ff q, \omega) $  dynamical structure factors is shown in
Fig. \ref{DCA_BSS.fig}  at $\beta t = 3$, $\delta \approx 14 \; \%$ and $U/t
=8$.  The BSS results correspond to
simulations on an  $8 \times 8$ lattice.
Fig. \ref{DCA_BSS.fig}b) depicts the BSS-QMC data in the spin sector.  Due to short-range
spin-spin correlations, remnants
of the spin-density-wave are observable, displaying a characteristic energy-scale of $2J$, where
$J$ is the usual exchange coupling, i.e. $J=4\frac{t^2}{U}$. The two-particle DCA calculations 
show spin excitations with the dominant weight concentrated, as
expected and seen in the QMC data, 
around the AF wavevector $(\pi,\pi)$. As apparent from the sum-rule, 
\begin{equation}
	\langle S^z(\ff q) S^z(- \ff q) \rangle = \frac{1}{\pi} \int {\rm d} \omega \; S(\ff q, \omega) 
\end{equation}
(see Fig. \ref{static.spin.fig} a)), the DCA overestimates the weight at this wave vector but 
does very well away from $\ff q = (\pi,\pi)$.  The dispersion in the two-particle data has again a higher 
energy branch around $2J$, but it also shows features at $J$. Since the total spin
is a conserved quantity, one expects a zero-energy excitation at $\ff q = (0,0)$. 
This is exactly reproduced in the $8 \times 8$ QMC-BSS data,  and qualitatively in the
DCA results. 

\begin{figure}[t]
  \begin{minipage}[t]{0.475\columnwidth}
    \flushleft{a)}
    \vfill
    \centering
    \resizebox{1.0\columnwidth}{!}{\includegraphics{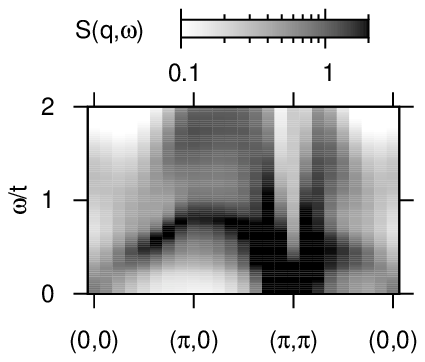}}
  \end{minipage}
  \hfill
  \begin{minipage}[t]{0.475\columnwidth}
    \flushleft{b)}
    \vfill
    \centering
    \resizebox{1.0\columnwidth}{!}{\includegraphics{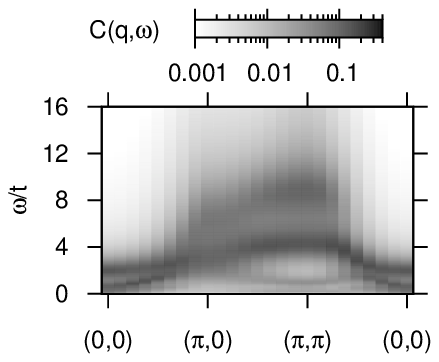}}
  \end{minipage}
  \begin{minipage}[t]{0.475\columnwidth}
    \flushleft{c)}
    \vfill
    \centering
    \resizebox{1.0\columnwidth}{!}{\includegraphics{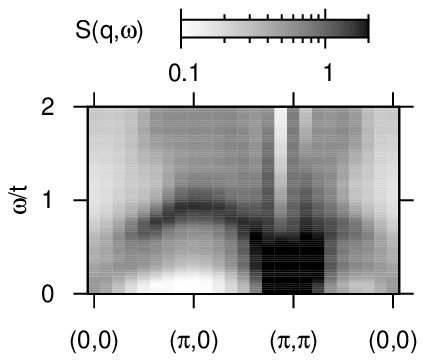}}
  \end{minipage}
  \hfill
  \begin{minipage}[t]{0.475\columnwidth}
    \flushleft{d)}
    \vfill
    \centering
    \resizebox{1.0\columnwidth}{!}{\includegraphics{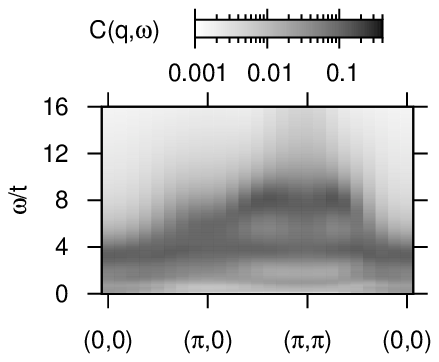}}
  \end{minipage}
  \begin{minipage}[t]{0.475\columnwidth}
    \flushleft{e)}
    \vfill
    \centering
    \resizebox{1.0\columnwidth}{!}{\includegraphics{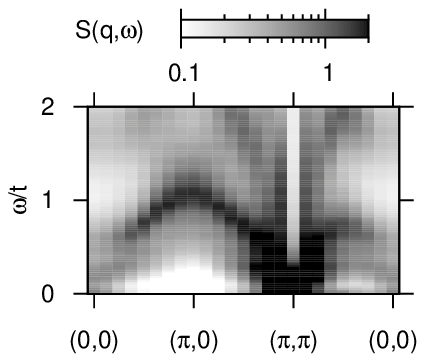}}
  \end{minipage}
  \hfill
  \begin{minipage}[t]{0.475\columnwidth}
    \flushleft{f)}
    \vfill
    \centering
    \resizebox{1.0\columnwidth}{!}{\includegraphics{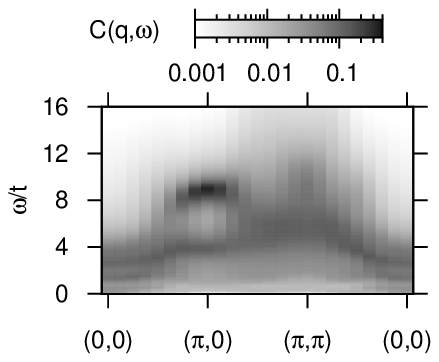}}
  \end{minipage}
  \begin{minipage}[t]{0.475\columnwidth}
    \flushleft{g)}
    \vfill
    \centering
    \resizebox{1.0\columnwidth}{!}{\includegraphics{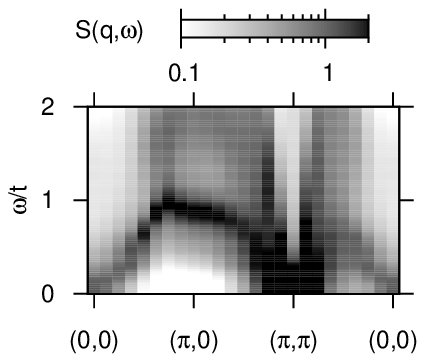}}
  \end{minipage}
  \hfill
  \begin{minipage}[t]{0.475\columnwidth}
    \flushleft{h)}
    \vfill
    \centering
    \resizebox{1.0\columnwidth}{!}{\includegraphics{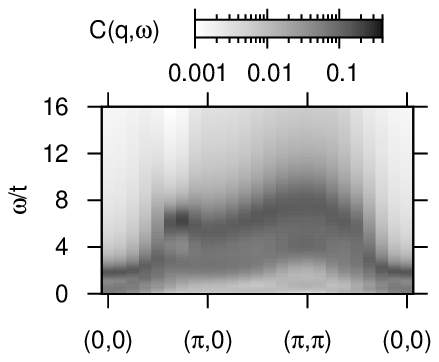}}
  \end{minipage}
  \caption{Dynamical spin and charge structure factors of the Hubbard
    model at $\beta t = 6$, $\delta \approx 14 \; \%$ and
    $U/t=8$. for different cluster sizes: (a-b): $L_c = 4$, (c-d):
    $L_c = 8$, (e-f): $L_c = 10$ and (g-h): $L_c = 16$. }
  \label{DCA_cluster.fig}   
\end{figure}

As a function of decreasing temperature,  the DCA dynamical spin structure factor shows a more 
pronounced  spin-wave spectrum.  This is confirmed in Fig. \ref{DCA_cluster.fig} on the left hand  side. 
Here, we fix the temperature to  $\beta t = 6$ and keep the doping at $\delta \approx 14 \; \%$  but vary the 
cluster size. As apparent, for all considered
cluster sizes ($L_c = 4,8,10,16$) a spin wave feature is indeed observable: a
peak maximum at $\ff q = (\pi,\pi)$ is present and the correct energy scale
at $\ff q = (\pi,0)$ with $2J$ is recovered. Unfortunately, a direct
comparison of both calculations at lower temperature is not possible
due to the severe minus-sign problem in the BSS calculation.

The investigation of the dynamical charge correlation function for the
above parameters shows that the 
DCA calculations, which are depicted in Fig. \ref{DCA_BSS.fig} c), can
also reproduce basic characteristics of the BSS
charge excitation spectrum \ref{DCA_BSS.fig} d). 
\begin{table}[b]
\caption{Values of $\alpha$ for the spectra in Fig. \ref{DCA_cluster.fig}.}
\begin{tabular}[b]{c|cccc}
\hline
\hline
Fig. \ref{DCA_cluster.fig} &\hspace{0.75cm}  a,b&\hspace{0.75cm} c,d&\hspace{0.75cm} e,f&\hspace{0.75cm} g,h\\
\hline
\hline
$\alpha$ (spin)&\hspace{0.75cm} 0.99&\hspace{0.75cm} 0.92&\hspace{0.75cm} 0.93&\hspace{0.75cm} 0.97 \\
$\alpha$ (charge)&\hspace{0.75cm} 0.98&\hspace{0.75cm} 1.00&\hspace{0.75cm} 1.00&\hspace{0.75cm} 1.00 \\
\hline
\hline
\end{tabular}
\label{alpha_tab}
\end{table}
Both calculations show excitations at $\omega \approx U  $ which
are set by the remnants of the Mott-Hubbard gap. Similar results are obtained at
lower temperatures ($\beta t = 6$) on the right hand side of
Fig. \ref{DCA_cluster.fig} for different cluster sizes ($L_c =
4,8,10,16$). The corresponding values of $\alpha$ are listed in
Tab. \ref{alpha_tab}. These values confirm the overall correctness of our approach 
in that the corresponding sum rule for the charge response is accurately (exactly 
for $\alpha = 1$) fulfilled.

\begin{figure}[t]
  \begin{minipage}[t]{0.475\columnwidth}
    \flushleft{a)}
    \vfill
    \centering
    \resizebox{1.0\columnwidth}{!}{\includegraphics{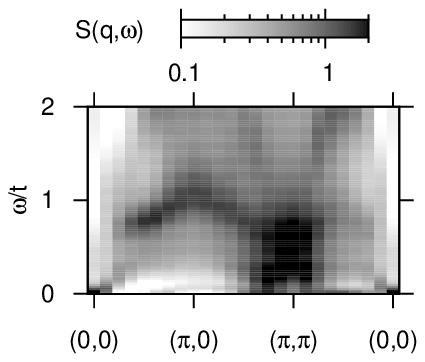}}
  \end{minipage}
  \hfill
  \begin{minipage}[t]{0.475\columnwidth}
    \flushleft{b)}
    \vfill
    \centering
    \resizebox{1.0\columnwidth}{!}{\includegraphics{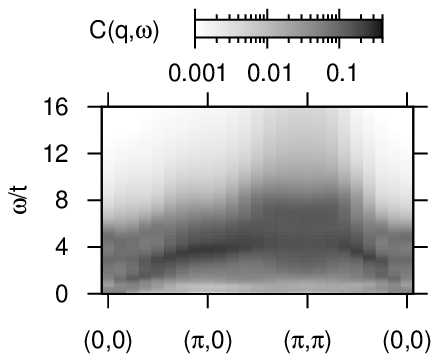}}
  \end{minipage}
  \begin{minipage}[t]{0.475\columnwidth}
    \flushleft{c)}
    \vfill
    \centering
    \resizebox{1.0\columnwidth}{!}{\includegraphics{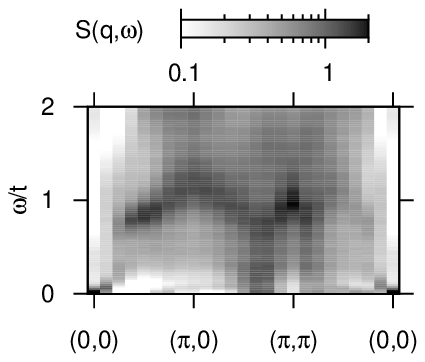}}
  \end{minipage}
  \hfill
  \begin{minipage}[t]{0.475\columnwidth}
    \flushleft{d)}
    \vfill
    \centering
    \resizebox{1.0\columnwidth}{!}{\includegraphics{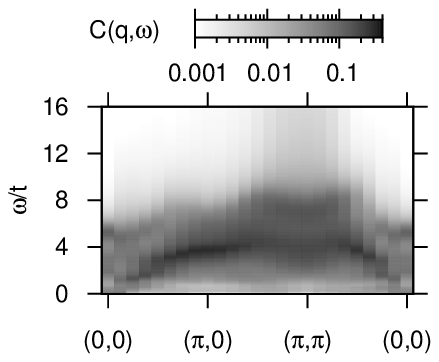}}
  \end{minipage}
  \caption{Spin- and charge structure functions for different
    dopings: (a,b): 27 \% and (c,d): 32 \%
    The calculations are carried out on an $L_c = 8 $ cluster at $\beta t =
    6$.}
  \label{sc_dop.fig}   
\end{figure}

\begin{figure}[t]
\begin{minipage}[t]{\columnwidth}
  \flushleft{a)}
  \vfill
  \centering
  \resizebox{0.85\columnwidth}{!}{\includegraphics{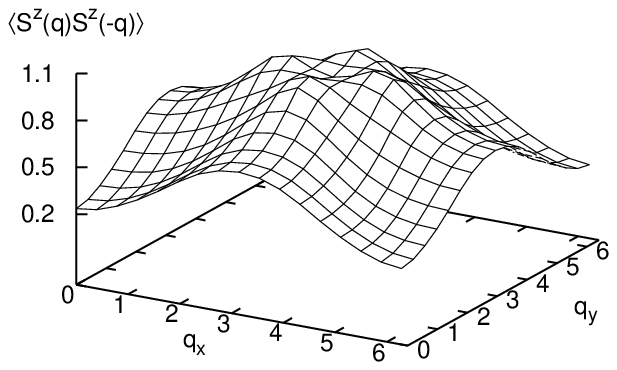}}
\end{minipage}
\hfill
\begin{minipage}[t]{\columnwidth}
  \flushleft{b)}
  \vfill
  \centering
  \resizebox{0.85\columnwidth}{!}{\includegraphics{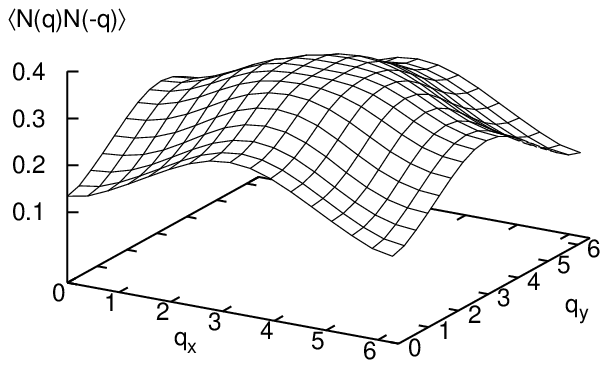}}
\end{minipage}
\caption{Static spin a) and charge b) correlation function at 
   $U/t=8$, $\delta \approx 32 \; \%$ and $\beta t = 6$ on an
   $L_c=8$ cluster.
}
\label{dca.s.c.static.fig} 
\end{figure}

The doping dependence of the spin- and charge-response is
examined in Fig \ref{sc_dop.fig}. Here, we restrict our
calculations to the $L_c=8$ cluster at $\beta t = 6$ and dopings between
$\delta = 14  \; \%$ and $\delta = 32 \; \%$.  
At $\delta = 14 \; \%$ (see Fig. \ref{DCA_cluster.fig}) the dynamical spin structure
factor displays a spin wave dispersion with  energy
scale $J$. That is  $E^{SDW}(\pi,0) = 2J$ with $J=4\frac{t^2}{U}$. As
the system is further doped ($\delta = 27 \; \%$) the dispersion
is no longer sharply peaked around $\ff q = (\pi,\pi)$. The excitations
broaden up and change their energy scale from $J=4\frac{t^2}{U}$ to
an energy scale set by the non-interacting bandwidth.  This effect becomes even more 
visible with higher dopings at  $\delta = 32 \; \%$ (Fig. \ref{sc_dop.fig} c)). 
Furthermore, the spectrum of the charge response shows a reduction of
the weight of states at high energies ($\omega/t \approx
8$). This behavior corresponds to the loss of weight of the upper
Hubbard band with increasing doping. 
The corresponding equal time spin and
charge correlation functions of Fig.~\ref{sc_dop.fig} c-d) are depicted in
Fig.~\ref{dca.s.c.static.fig}. As in  auxiliary-field QMC simulations \cite{IFT98}, 
the equal time spin correlation function  shows a set of peaks 
 at $\ff q = (\pi \pm \epsilon  ,\pi)$ and $\ff q = (\pi   ,\pi \pm \epsilon)$.
Here $\epsilon $ is proportional to the doping. 
The overall trend of the doping dependence of the spin- and
charge-responses is in good agreement with the previous findings of
QMC simulations (\oocite{preuss95,preuss97}): there it was shown that the
spin-response  has a characteristic energy
scale $\omega \approx 2J$ and an SDW-like dispersion up to about
$\delta \approx 10 - 15 $ \% doping, despite the fact that at these
dopings  the spin-spin
correlations are very short-ranged (of order of the lattice
parameter).

\begin{figure}[t]
  \begin{minipage}[t]{\columnwidth}
    \flushleft{a)}
    \vfill
    \centering
    \resizebox{0.65\columnwidth}{!}{\includegraphics{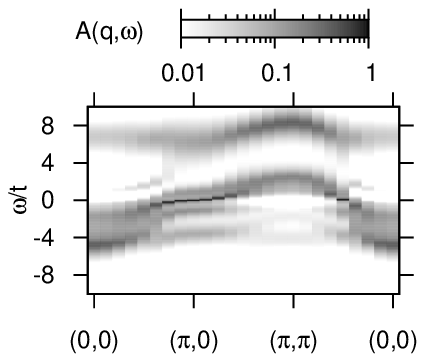}}
  \end{minipage}
  \hfill
  \begin{minipage}[t]{\columnwidth}
    \flushleft{b)}
    \vfill
    \centering
    \resizebox{0.65\columnwidth}{!}{\includegraphics{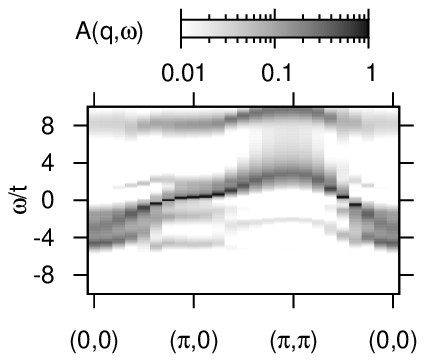}}
  \end{minipage}
  \begin{minipage}[t]{\columnwidth}
    \flushleft{c)}
    \vfill
    \centering
    \resizebox{0.65\columnwidth}{!}{\includegraphics{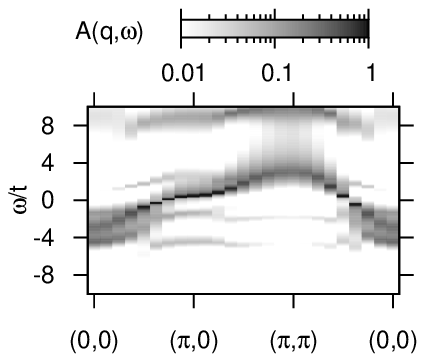}}
  \end{minipage}
  \caption{Angle-resolved spectral functions $A(\ff q,\omega)$ for
    various hole dopings: a): 14 \%, b): 27 \% and c): 32 \%. 
    Calculations are carried out on an $L_c = 8 $ cluster at $\beta t = 6$.}
  \label{akom_dop.fig}   
\end{figure}

A lot of the features of the two-particle spectra have direct
influence on the single-particle spectral function and vice-versa. 
At optimal doping, $\delta = 14 $ \% the spectral function $A(\ff q,\omega)$ in
Fig. \ref{akom_dop.fig} a) shows three distinguish features. 
An upper Hubbard band ($\omega/t \approx 8t$) and  a lower
Hubbard band which splits in an incoherent  background and a
quasiparticle band of width set by the magnetic scale $J$. 
In agreement with earlier QMC data (\oocite{preuss95,preuss97}), we view this narrow
quasiparticle  band as a fingerprint of a spin polaron
where the bare particle is dressed by spin fluctuations. 
The fact that the dynamical spin structure factor in Fig
\ref{DCA_cluster.fig} c)
shows a well  defined magnon despersion at this temperature and doping,  $ \delta =
14 $ \%, allows us to interpret the features centered around  $ \ff q = (0,0) $
and below the  Fermi energy  as backfolding  or shadows of the
quasiparticle  band at $\ff q  = (\pi,\pi)$.
A comparison of the charge response spectrum in 
Fig. \ref{DCA_cluster.fig} d) with the corresponding  single-particle
spectra in Fig. \ref{akom_dop.fig} a) reveals that
the response in the particle-hole channel at almost
zero energy is caused by particle-hole excitations around the
quasi-particle spin-polaron band 
close to the Fermi energy. The high energy excitations,
mentioned above, are due to transitions from the quasi-particle band
to the upper Hubbard band. 

As a function of doping, notable changes in the spectral function
which  are reflected in the two-particle properties are apparent. On one
hand, the spectral weight in the upper Hubbard band is reduced. As
mentioned previously, this reduction  in high energy spectral weight
is apparent in the  dynamical charge structure factor. 
On the other hand, at higher dopings the magnetic fluctuations are
suppressed. Consequently, the narrow band changes its bandwidth from
the magnetic exchange energy $J$ to the free bandwidth.
This evolution is clearly apparent in
Figs. \ref{akom_dop.fig} b) and c) and is in good agreement with
previous BSS-QMC results~\cite{preuss97}. 

\section{Conclusions}
\label{sec:con}
Two-particle spectral functions, such as spin- and charge- dynamical spin structure factors, are clearly 
quantities  which are of crucial relevance for the understanding of correlated materials. Within the DCA, those 
quantities have remained elusive due to numerical complexity.   In principle, the two-particle irreducible vertex, 
$\Gamma_{\underline{K}',\underline{K}''} (\underline{Q}) $, containing three momenta and frequencies has to  be 
extracted from the cluster and  inserted in the Bethe-Salpeter equation.   Calculating  this quantity on 
the cluster is in principle possible, but corresponds to a daunting task which  --- to the best of our knowledge --- 
has never been carried out.  To circumvent this problem,  we have proposed a simplification which relies on 
the assumption that $\Gamma_{\underline{K}',\underline{K}''} (\underline{Q}) $ is only weakly dependent on 
$\underline{K}'$ and $\underline{K}''$. Given the validity of this assumption, one can average over 
$\underline{K}'$ and $\underline{K}''$  and retain  its dependency on
frequency and  momenta,  $\underline{Q}$,  of the 
excitation.  We have tested this idea extensively for the two dimensional Hubbard model  at strong  couplings, 
$U/t = 8$.   At dopings $\delta \approx 10 $ \% and higher, we have  found a  good agreement  between the
N\'eel temperature on an $L_c=8$ lattice as calculated within a
symmetry broken DCA scheme with our new two-particle approach.   This finding 
lends support to the validity of our scheme in this doping range.  Furthermore, we studied  the doping and temperature 
dependence of the spin and charge dynamical  structure factors as well as the single-particle spectral function at $\beta t = 6$. 
Our  results provide  a  consistent picture of the physics of doped Mott
insulators, very reminiscent of previous findings  within  auxiliary 
field QMC simulations. The strong point of the method, in contrast to auxiliary field  QMC approaches, is that 
it can be pushed to lower temperatures  above and below the superconducting transition temperature. Work in 
this direction is presently under progress. 
\\

\acknowledgments
Discussions with D.J. Scalapino about many ideas of this approach in the early 
stages are gratefully acknow\-ledge. 
This work has been supported by the Deutsche
Forschungsgemeinschaft within the Forschergruppe FOR~538,
by the Leibniz Rechenzentrum Munich for providing computer
resources, in part by
the National Science Foundation under Grant No. PHY 05-51164,
and as well as by KONWIHR. We thank M. Potthoff, E. Arrigoni, L. Martin, S. Brehm
and M. Aichhorn for conversations.

\end{document}